\documentclass[aps,prd,preprint,groupedaddress,preprintnumbers,showpacs]{revtex4-1}
\usepackage{graphicx}
\newcommand{\e}{\hbox{e}}
\newcommand{\ep}{\epsilon}
\newcommand{\bra}[1]{\left\langle #1 \right|}
\newcommand{\ket}[1]{\left| #1 \right\rangle}

\newcommand{\lb}{\left\lbrace}
\newcommand{\rb}{\right\rbrace}
\begin{document}

\preprint{KOBE-TH-13-10}

\title{Renormalization for free harmonic oscillators}


\author{H.~Sonoda}
\email[]{hsonoda@kobe-u.ac.jp}
%\thanks{}
\affiliation{Physics Department, Kobe University, Kobe 657-8501 Japan}


\date{1 February 2014}

\begin{abstract}
    We introduce a model of free harmonic oscillators that requires
    renormalization.  The model is similar to but simpler than the
    soluble Lee model.  We introduce two concrete examples: the first,
    resembling the three dimensional $\phi^4$ theory, needs only mass
    renormalization, and the second, resembling the four dimensional
    $\phi^4$ theory and the Lee model, needs additional
    renormalization of a coupling and a wave function.
\end{abstract}


\maketitle

%\section{}
%\subsection{}
%\subsubsection{}

% If in two-column mode, this environment will change to single-column
% format so that long equations can be displayed. Use
%\begin{widetext}
%\end{widetext}

\section{Introduction}

The purpose of this short paper is to introduce simple examples of
renormalization using a model of free harmonic oscillators.  The model
was originally introduced by Dirac \cite{Dirac:book} for his
explanation of resonance scattering.  We have simplified Dirac's model
slightly by transcribing it in terms of harmonic oscillators.

We consider two concrete examples.  The first example requires
renormalization of only a frequency, and it resembles the $\phi^4$
theory in three dimensions.  The second example requires additional
renormalization of a dimensionless coupling, and it resembles the
$\phi^4$ theory in four dimensions.  The latter example is
particularly illuminating since it suffers from the Landau pole just
as the four dimensional $\phi^4$ theory.  

Since our model is free, we can compute its Green function by summing
a geometric series.  In this sense the model captures the essence of
the soluble Lee model \cite{Lee:1954iq} and the large $N$ limit of the
O($N$) linear sigma model.  Our model is simpler thanks to the use of
harmonic oscillators.

The implication of the Landau pole for our second example is exactly
the same as for the four dimensional $\phi^4$ theory.  To keep a
non-vanishing interaction, we must keep the ultraviolet cutoff of the
theory finite.  We will derive the exact cutoff dependence of the
renormalized coupling as for the Lee model and the large $N$ limit of
the O($N$) linear sigma model.

In Appendix \ref{appendix-Lee} we show that part of the Lee model can
be reproduced exactly with a judicious choice of frequency dependence
of the coupling in our free model.

\section{The model}

We consider the hamiltonian for a collection of harmonic oscillators:
\begin{equation}
H = H_F + H_I
\end{equation}
where
\begin{equation}
\lb\begin{array}{c@{~=~}l}
H_F & \Omega a^\dagger a + \sum_n \omega_n a_n^\dagger a_n\\
H_I & - \sum_n g_n (a_n^\dagger a + a^\dagger a_n)
\end{array}\right.
\end{equation}
This is a model of coupled oscillators.  Since the hamiltonian is
quadratic in oscillators, the model is free.  But a model as simple as
this can be interesting and useful, since it admits a variety of
interpretations.  Here are three examples:
\begin{enumerate}
\item $a$ stands for a charged oscillator, and $a_n$ for modes of
  radiation.  Thus, the model mimics an atom unstable under a
  radiative decay.
\item $a$ stands for a meson in its center of mass system, and $a_n$
  for a pair of decay products whose relative momentum is oriented in
  a direction denoted by $n$.
\item $a$ stands for a mode mediating an attractive force between a
    Cooper pair of electrons denoted by $n$.  (The model of Cooper
    \cite{Cooper:1956} is reproduced in the limit $\Omega \to \infty$,
    where $g_n^2/\Omega$ is a fixed frequency. See Appendix
    \ref{appendix-Cooper}.)
\end{enumerate}

The model reduces to the Jaynes-Cummings model \cite{Jaynes:1963} if we
single out a particular mode $n$.

\section{Green function}

Let us consider a complex valued Green function defined by
\begin{equation}
G (z) \equiv \bra{0} a \frac{1}{z - H} a^\dagger \ket{0}
\end{equation}
We can sum the geometric series given by perturbation theory as
\begin{equation}
G (z) = \frac{1}{z - \Omega} \sum_{L=0}^\infty \left(\sum_n g_n^2
\frac{1}{z - \omega_n} \frac{1}{z - \Omega}\right)^L
= \frac{1}{z - \Omega - \sum_n g_n^2 \frac{1}{z - \omega_n}}
\end{equation}
If we denote the physical size (or ``volume'') of the system by $V$,
the number of modes $n$ contained in a finite frequency interval is
proportional to $V$.  Hence, we can define the density of states per
unit volume by
\begin{equation}
\frac{dn}{d\omega} \equiv \lim_{V\to\infty} \frac{1}{V} \sum_n \delta (\omega-\omega_n) 
\end{equation}
Assuming that $g_n^2$ is of order $\frac{1}{V}$, we obtain a finite
non-negative function
\begin{equation}
g_\omega^2 \equiv \lim_{V \to \infty} \sum_n g_n^2 \,\delta (\omega -
\omega_n)
\end{equation}
in this thermodynamic limit.  Note $g_\omega^2$ has the dimension of a
frequency.  Hence, in the limit we obtain
\begin{equation}
G (z) = \frac{1}{z - \Omega - \int d\omega\, g_\omega^2 \frac{1}{z -
    \omega}}
\end{equation}

Let us suppose $g_\omega^2$ is non-vanishing only in a finite range $[
\omega_L, \omega_H ]$ of $\omega$.  For example, if $a_n$ denotes a
pair of electrons, we may take $\omega_L$ to be twice the electron
mass. The choice of the band width $\omega_H - \omega_L$ depends on
the model.  It may be a finite Debye temperature as in Cooper's model,
or we may wish to take $\omega_H$ to infinity as in the case of a
meson decay.

For $\omega \in [\omega_L, \omega_H]$, we obtain the imaginary part:
\begin{equation}
\Im G (\omega + i \ep) = \frac{- \pi g_\omega^2}{b_\omega^2
+ \pi^2 g_\omega^4}
\end{equation}
where
\begin{equation}
b_\omega \equiv \Re G (\omega + i \ep)^{-1} = \omega - \Omega -
\int_{\omega_L}^{\omega_H} d\omega'\, g_{\omega'}^2 \mathbf{P}
\frac{1}{\omega - \omega'}
\end{equation}
Hence, we obtain the dispersion relation
\begin{equation}
G (z) = \sum_i \frac{r_i}{z - \omega_i} + \int_{\omega_L}^{\omega_H}
d\omega \frac{1}{z - \omega} \,\rho (\omega)
\end{equation}
where $\omega_i$ are isolated poles with positive residues $r_i$, and
the spectral function is defined by
\begin{equation}
\rho (\omega) \equiv \frac{1}{\pi} (-) \Im G (\omega + i \ep) =
\frac{g_\omega^2}{b_\omega^2 + \pi^2 g_\omega^4}\quad
(\omega_L < \omega < \omega_H)
\end{equation}
The asymptotic behavior
\begin{equation}
G (z) \stackrel{|z|\to\infty}{\longrightarrow} \frac{1}{z}
\end{equation}
gives the sum rule
\begin{equation}
\sum_i r_i + \int_{\omega_L}^{\omega_H} d\omega\, \rho (\omega) = 1
\end{equation}

\section{First example}

\begin{figure}[h]
\centering
\includegraphics[width=8cm]{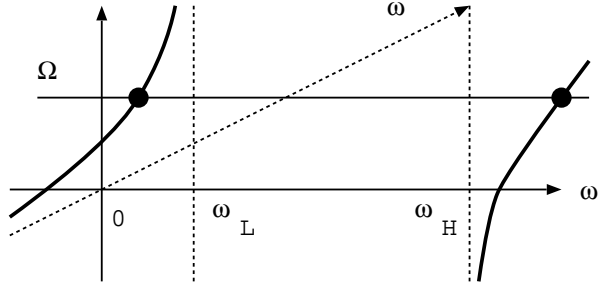}
\caption{The dark curves give $\omega - g^2 \ln
  \frac{\omega-\omega_L}{\omega-\omega_H}$, which equals $\Omega$ for
  a bound state.}
\label{example1}
\end{figure}

The first example is given by
\begin{equation}
g_\omega^2 = g^2 > 0\quad (\omega_L < \omega < \omega_H)
\end{equation}
where $g^2$ is a constant frequency.  Since
\begin{equation}
\int_{\omega_L}^{\omega_H} d\omega\, \frac{1}{z - \omega} = \ln
\frac{z - \omega_L}{z - \omega_H}
\end{equation}
we obtain
\begin{equation}
G(z)^{-1} = z - \Omega - g^2 \ln \frac{z - \omega_L}{z - \omega_H}
\end{equation}
By plotting $\omega - g^2 \ln (\omega-\omega_L)/(\omega -\omega_H)$
for $\omega < \omega_L$ and $\omega > \omega_H$ (Fig.~\ref{example1}),
we find two isolated states, one below $\omega_L$ and another above
$\omega_H$.  Hence, the force mediated by the $\Omega$ mode is
attractive for the state below $\omega_L$, and repulsive for that
above $\omega_H$.
This is easy to understand.  The second order perturbation theory gives
the correction to the energy of the mode $\omega_n$ as
\[
\Delta \omega_n = g_n^2 \frac{1}{\omega_n - \Omega}
\]
This is negative for $\omega_n < \Omega$, and positive for $\omega_n >
\Omega$.

We now take $\omega_H$ large.  We then obtain
\begin{equation}
G(z)^{-1} = z - \Omega - g^2 \ln \frac{\omega_L - z}{\mu} + g^2 \ln
\frac{\omega_H}{\mu}
\end{equation}
By defining a renormalized frequency
\begin{equation}
\Omega_r \equiv \Omega - g^2 \ln \frac{\omega_H}{\mu}
\end{equation}
where $\mu$ is a renormalization scale, we obtain the renormalized
Green function as
\begin{equation}
G_r (z) \equiv \lim_{\omega_H \to \infty} G(z) = \frac{1}{z - \Omega_r
- g^2 \ln \frac{\omega_L - z}{\mu} }
\end{equation}
This satisfies the renormalization group equation
\begin{equation}
\left( \mu \frac{\partial}{\partial \mu} + g^2
    \frac{\partial}{\partial \Omega_r} \right) G_r (z) = 0
\end{equation}
$G_r$ has only one pole at $\omega = \omega_b$, which satisfies
\begin{equation}
\omega_b - \Omega_r - g^2 \ln \frac{\omega_L - \omega_b}{\mu} = 0
\end{equation}
This is solved explicitly as
\begin{equation}
\omega_b = \omega_L - g^2 W_0 \left( \frac{\mu}{g^2}\, \e^{-
      \frac{\Omega_r}{g^2}} \right)
\end{equation}
where $W_0 (x)$ is the main branch of the Lambert W
function\cite{Corless:1996}, satisfying 
\begin{equation}
W_0 (x) \exp \left( W_0 (x) \right) = x
\end{equation}

The dispersion relation for the renormalized Green function is given
by
\begin{equation}
G_r (z) = \frac{r_b}{z - \omega_b} + \int_{\omega_L}^\infty d\omega\,
\frac{1}{z - \omega} \,\rho (\omega)
\end{equation}
where
\begin{equation}
\rho (\omega) \equiv \frac{g^2}{\left( \omega - \Omega_r - g^2 \ln
      \frac{\omega - \omega_L}{\mu} \right)^2 + \pi^2 g^4} 
\end{equation}
For $g^2 \ll \Omega_r$, we find $r_b \ll 1$, and the spectral function
$\rho (\omega)$ is sharply peaked at $\Omega_r$ with width $\pi g^2$.
(In fact there is an additional peak of an extremely narrow width $\mu
\e^{- \frac{\Omega_r - \omega_L}{g^2}}$ just above $\omega_L$.)  We
obtain the approximate sum rule
\begin{equation}
\int_{\omega_L}^\infty d\omega\,\rho (\omega) \simeq 1
\end{equation}

\section{Second example}

The second example is given by
\begin{equation}
g_\omega^2 = \omega\,\bar{g}^2 \quad (\omega_L < \omega < \omega_H)
\end{equation}
where $\bar{g}^2$ is a dimensionless positive constant.  This model
has a stronger coupling toward the high frequencies.

\begin{figure}[h]
\centering
\includegraphics[width=10cm]{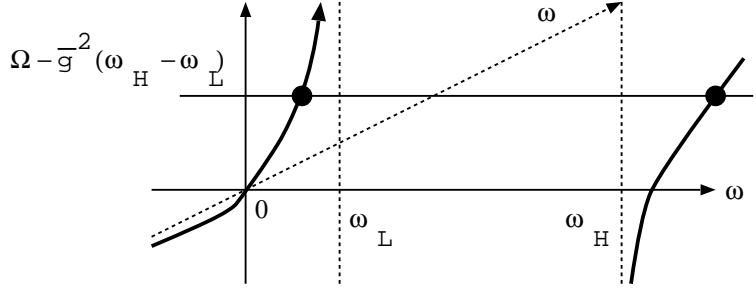}
\caption{The dark curves give $\omega \left(1 - \bar{g}^2 \ln
      \frac{\omega_L-\omega}{\omega_H-\omega} \right)$ for $\omega <
  \omega_L$ and $\omega > \omega_H$}
\label{example2}
\end{figure}

The Green function is obtained as
\begin{equation}
G(z)^{-1} = z - \Omega + \bar{g}^2 \lb \omega_H - \omega_L - z \ln
\frac{z - \omega_L}{z - \omega_H} \rb
\end{equation}
As in the first example, there are two isolated states, one below
$\omega_L$ and another above $\omega_H$.  (Fig.~\ref{example2})

Let us now consider the limit $\omega_H \to \infty$.  To get a limit,
we must renormalize not only $\Omega$ but also $\bar{g}^2$.  We define
renormalized parameters by
\begin{eqnarray}
\Omega_r &\equiv& \frac{1}{Z}
\left( \Omega - \bar{g}^2 (\omega_H - \omega_L) \right)\\
\bar{g}_r^2 &\equiv& \frac{\bar{g}^2}{Z}\label{g2r}
\end{eqnarray}
where
\begin{equation}
Z \equiv 1 + \bar{g}^2 \ln \frac{\omega_H}{\mu}
\end{equation}
We then obtain the renormalized Green function as
\begin{equation}
G_r (z) \equiv \lim_{\omega_H \to \infty} Z \cdot G (z) =
\frac{1}{z - \Omega_r - \bar{g}_r^2 z \ln \frac{\omega_L - z}{\mu}}
\end{equation}
Note the necessity of a wave function renormalization by the factor
$Z$.  The renormalized Green function satisfies the renormalization
group equation
\begin{equation}
\left( \mu \frac{\partial}{\partial \mu} + \bar{g}_r^2 \Omega_r
  \frac{\partial}{\partial \Omega_r} + \bar{g}_r^4
    \frac{\partial}{\partial \bar{g}_r^2} \right) G_r (z) = -
\bar{g}_r^2\, G_r (z)
\end{equation}
implying the anomalous dimension $\bar{g}_r^2$ of the Green function.

\begin{figure}[h]
\centering
\includegraphics[width=10cm]{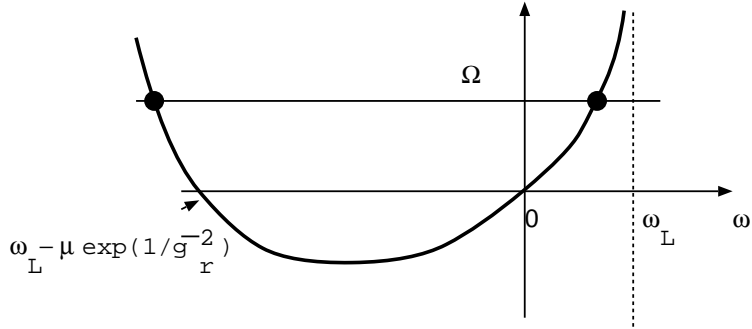}
\caption{The dark curve gives $\omega \left(1 - \bar{g}_r^2 \ln
  \frac{\omega_L - \omega}{\mu}\right)$.  The tachyon pole lies below
  $\omega_L - \mu \,\e^{1/\bar{g}_r^2}$}
\label{example2-renormalized}
\end{figure}

Contrary to our expectation that the renormalized Green function has
only one pole just below $\omega_L$, we find an additional pole
$\omega_t$ which is very negative.  (Fig.~\ref{example2-renormalized})
We call this a tachyon since the residue $r_t$ of the pole at
$\omega_t$ is negative:
\begin{equation}
G_r (z) \stackrel{z \to \omega_t}{\longrightarrow} \frac{r_t}{z -
  \omega_t}\quad (r_t < 0)
\end{equation}
This tachyon pole is reminiscent of the tachyon pole in the large $N$
limit of the O($N$) linear sigma model in four
dimensions.\cite{Coleman:1974jh}

The tachyon pole arises since we cannot really take $\omega_H$ all the
way to infinity.  In the limit $\omega_H \to \infty$, we get a trivial
result:
\begin{equation}
\bar{g}_r^2 = \frac{1}{\frac{1}{\bar{g}^2} + \ln \frac{\omega_H}{\mu}}
\stackrel{\omega_H \to \infty}{\longrightarrow} 0
\end{equation}
To find the largest possible $\omega_H$, we use $1/\bar{g}^2 \ge 0$ to
obtain
\begin{equation}
\bar{g}_r^2 \le \frac{1}{\ln \frac{\omega_H}{\mu}}
\end{equation}
Hence,
\begin{equation}
\frac{\omega_H}{\mu} \le \exp \left( \frac{1}{\bar{g}_r^2}\right)
\end{equation}
The equality corresponds to the Landau pole
\begin{equation}
\bar{g}^2 = +\infty
\end{equation}
Thus, to be rid of the tachyon, we must keep $\omega_H$ large but
finite.  The same resolution works for the Lee model and the large $N$
limit of the four dimensional scalar theory.

\appendix

\section{Renormalized Cooper's model\label{appendix-Cooper}}

We consider the strong coupling limit of the first example.  Let
\begin{equation}
g^2 = \bar{g}^2 \,\Omega_r
\end{equation}
and take the limit $\Omega_r \to \infty$.  We obtain
\begin{equation}
\lim_{\Omega_r \to \infty} \frac{1}{\Omega_r} G_r (z)^{-1} = - 1 -
\bar{g}^2 \ln \frac{\omega_L - z}{\mu}
\end{equation}
Hence, the bound state energy is given by
\begin{equation}
\omega_L - \omega_b = \mu \exp \left( - \frac{1}{\bar{g}^2} \right)
\end{equation}

\section{Lee's model\label{appendix-Lee}}

The Lee model \cite{Lee:1954iq} is a non-relativistic model describing
an interaction of a fermion with a meson of mass $m_\theta$.  The
fermion comes in two flavors: a V-particle with mass $m_V$ and an
N-particle with mass $m_N$.  They only interact via
\[
V \longleftrightarrow N + \theta
\]
so that the number of V-particles plus N-particles, $N_V + N_N$, and
the number of V-particles plus $\theta$-mesons, $N_V + N_\theta$, are
conserved.  Using our free field model, we can reproduce exactly the
spectrum of the states satisfying $N_V + N_N = 1$ \& $N_V + N_\theta =
1$ (one V or a pair of N and $\theta$).  The V-particle corresponds to
the mode $\Omega = m_V - m_N + \delta m_V$, and the pair of N and
$\theta$ of momentum $k$ corresponds to the mode $\omega_n = \sqrt{k^2
  + m_\theta^2}$.  $m_\theta$ plays the role of our $\omega_L$.  The
hamiltonian of this subspace is identical with that of our free model
with the choice
\begin{equation}
g_\omega^2 = \frac{g^2}{4 \pi^2} \sqrt{\omega^2 - m_\theta^2}
\end{equation}
For large $\omega \gg m_\theta$, we find $g_\omega^2 \propto \omega$,
and our second model shares the same renormalization properties as the
Lee model.

In our notation Eq. (8) of Lee is given by
\begin{equation}
\delta m_V = - \int_{m_\theta}^{\omega_H} d\omega\, \frac{g_\omega^2}{m_V - m_N -  \omega}
\end{equation}
and Eq. (10) of Lee for the renormalization constant is given by
\begin{equation}
Z_2^{-1} = 1 + \int_{m_\theta}^{\omega_H} d\omega \frac{g_\omega^2}{\left( m_V - m_N - \omega\right)^2}
\end{equation}
We note that $\delta m_V$ is linearly divergent, and that $Z_2^{-1}$ is logarithmically divergent, as $\omega_H \to \infty$.

Finally, the phase shift $\delta$, defined by the phase of the Green function $G (\omega + i \ep)$, is given by
\begin{equation}
\tan \delta \equiv \frac{\pi g_\omega^2}{- b_\omega}
= \frac{\pi g_\omega^2}{m_V - m_N - \omega} \left( 1 + \int_{\omega'} \frac{g_{\omega'}^2}{m_V - m_N - \omega'} \mathbf{P} \frac{1}{\omega - \omega'}\right)^{-1}
\end{equation}
which agrees with Lee's (16).  ($\omega_0$ is replaced by $\omega$ here.)

\begin{acknowledgments}
    I thank Yuji Igarashi \& Katsumi Itoh for discussions.  I also
    thank the referee for the comments that improved the clarity of
    the manuscript.  This work was partially supported by the JSPS
    grant-in-aid \# 25400258.
\end{acknowledgments}

\bibliography{free-field}

\end{document}